\documentclass[12pt,epsfig]{article}
\usepackage{amsmath}
\usepackage{graphicx}
\usepackage{epstopdf}
\usepackage{lscape}
\usepackage{longtable}
\usepackage{float}
\usepackage[utf8]{inputenc}
\usepackage[usenames, dvipsnames]{color}
\begin{document}
	\begin{center}
		\textbf {A new interpoint distance-based clustering algorithm using kernel density estimation}\\
	\end{center}
	\begin{center}
		Dr. Soumita Modak$^{*}$\\
		Faculty of Statistics\\
		Department of Statistics\\
		University of Calcutta\\
		Basanti Devi College\\
		147B, Rash Behari Ave, Kolkata- 700029, India\\
		Email: soumitamodak2013@gmail.com\\
		Orcid id: 0000-0002-4919-143X\\
		Homepage: https://sites.google.com/view/soumitamodak
		
	\end{center}
	
	Abstract: A novel nonparametric clustering algorithm is proposed using the interpoint distances between the members of the data to reveal the inherent clustering structure existing in the given set of data, where we apply the classical nonparametric univariate kernel density estimation method to the interpoint distances to estimate the density around a data member. Our clustering algorithm is simple in its formation and easy to apply resulting in well-defined clusters. The algorithm starts with objective selection of the initial cluster representative and always converges independently of this choice. The method finds the number of clusters itself and can be used irrespective of the nature of underlying data by using an appropriate interpoint distance measure. The cluster analysis can be carried out in any dimensional space with viability to high-dimensional use. The distributions of the data or their interpoint distances are not required to be known due to the design of our procedure, except the assumption that the interpoint distances possess a density function. Data study shows its effectiveness and superiority over the widely used clustering algorithms.   
	
	keyword: Clustering algorithm, Interpoint distance, Nonparametric method, Kernel density estimator, High-dimensional applicability. 
	
	\section{Introduction}
	Cluster analysis is the unsupervised classification procedure to classify a set of data into homogeneous groups called clusters such that we obtain distinct meaningful classes where similar or closer data members are clustered in the same group and further data members in different groups. Cluster analysis is very much required for big data study, where clustered data reveal the different sources of data generation behind the clusters. The literature have diverse clustering methods like partitioning, hierarchical, model-based, grid-based and density-based (Ruspini 1970; Hartigan 1975; Hartigan and Wong 1979; Bezdek 1981; Ester et al. 1996; Jain et al. 1999; McLachlan and Peel 2000; Kaufman and Rousseeuw 2005; Campello et al. 2013; Arias-Castro et al. 2016; Matioli et al. 2018; Modak et al. 2018, 2020, 2022; Modak 2019; Tarnopolski 2019; Toth et al. 2019; Cheng et al. 2021; Modak 2021). We consider the hard-cluster analysis with all the data members classified into mutually exclusive and exhaustive clusters. Here we measure the closeness or similarity of two members of data through the interpoint distance computed between them.
	
	Based on the interpoint distances, widely applied clustering algorithms are available to perform an unsupervised classification (Hartigan and Wong 1979; Kaufman and Rousseeuw 2005; Campello et al. 2013), and there exist efficient clustering accuracy measures (Dunn 1974; Handl et al. 2005; Modak 2022a, b) which are used as cluster validity indices to determine the quality of a classification using the properties of the clusters gained through a clustering method. Our interpoint distance-based clustering algorithm works on data measured on arbitrary scales with the help of an appropriately selected distance measure (may not be strictly metric). Interpoint distance allows the method's use for univariate, multivariate or  high-dimensional data, where the number of observations can be close to or less than the number of variables under study. Our nonparametric algorithm does not make use of any parametric models or the distribution of the given data. The only assumption adopted here is that the data is possessing interpoint distances with a density function (may not be known). It finds clusters of a reasonable size (user-defined) around data members with a maximum density within the analyst's specified neighborhood, where the density around a member is determined through estimating the density function of its interpoint distances applying the classical nonparametric univariate kernel density estimator (Silverman 1986; Wand and Jones 1995; Bandyopadhyay and Modak 2018). 
	
	Our method is implemented using an algorithm which starts precisely and objectively determining the initial cluster representative in contrast to the  widely-used existing algorithms like $K-$means (MacQueen 1967; Hartigan and Wong 1979) wherein, as the initial choice of the cluster centers is not fixed, it can change the outcome and, therefore, to achieve acceptable results usually multiple repetitions are needed over different random initial choices, provided not all of them guarantee the algorithm's convergence; whereas $K-$medoids method using `PAM' algorithm (Kaufman and Rousseeuw 2005) calls for the computation of an additional phase involving iterative steps to finalize the initial choice of the cluster medoids. Our algorithm ensures convergence, irrespective of the selection for the initial cluster representative, producing well-defined clusters of a minimum size specified by the user-defined choice of $n'$ (positive integer), and another tuning parameter the algorithm depends on is $h$ $(>0)$ to form some neighborhood around a member, which is also used as a smoothing parameter in the kernel density estimator. The design of our algorithm makes it converge with any selected values for two of its hyperparameters.
	The proposed clustering method itself evaluates the unknown true value for the number of existing clusters $(K)$ in the data set rather than requiring it to be specified as a priori unlike other popular clustering algorithms like $K-$means, $K-$medoids or hierarchical clustering methods (Hartigan 1975; Hartigan and Wong 1979; Kaufman and Rousseeuw 2005), where the whole clustering has to be performed for different values of $K$ and then, finally choose a value corresponding to the best possible classification reached in terms of an efficient cluster accuracy measure. Thus, our method saves that computational burden. It is shown to outperform the popular density-based `DBSCAN' algorithm (Ester et al. 1996; Campello et al. 2013; Hahsler et al. 2019; Modak 2022a), efficient enough to expose arbitrary-shaped clusters unlike the previously mentioned competitors $K$-means or $K$-medoids, and another recent kernel-based algorithm `ClusterKDE' (Matioli et al. 2018), with $K$ not needed as a priori. However, DBSCAN has two tuning parameters whose values are crucial for the resulting clusters, and therefore, to be chosen carefully, which is another concern; while the existing automatic selection techniques (Ester et al. 1996; Hahsler et al. 2019) do not guarantee the best outcome, it needs subjective interference, that no doubt becomes time-consuming. Moreover, for the chosen values, some members may not belong to any clusters and thereby marked as noise, which needs further subjective analysis. On the other hand, the nonparametric classical kernel-based ClusterKDE algorithm has very limited applications for being useful only to a maximum of 2-dimensional data, where different values of its bandwidth parameter(s) do not always guarantee convergence of the algorithm. This approach suffers from the curse of dimensionality, i.e. with increasing dimension, the number of parameters involved in the algorithm increases. Anyway, the applicability of our advised method is demonstrated by (a) outlier affected data, with overlapping classes, measured on arbitrary scales (Kaufman and Rousseeuw 2005), (b) benchmark data set `Ruspini' (Ruspini 1970), (c) multivariate data simulated with complex dependence structure formed by a copula (Nelsen 2006; Modak and Bandyopadhyay 2019), (d) closely placed groups of arbitrary shapes with noisy observations, (e) real-life bivariate spatial sample (Matioli et al. 2018) and (d) high-dimensional biostatistical data (Alon et al. 1999). This data study shows its superior performance in comparison with the other methods from the literature.  
	
	The paper is designed as follows. Section 2 proposes our novel method and analyzes it in detail. Section 3 demonstrates its application through data study in terms of synthetic and real-life sets of data. Conclusions are drawn in Section 4.
	\section{Method}
	Let $M_1,\ldots,M_n$ be the members or the corresponding (univariate or multivariate) observations belonging to the set $R^p$ ($p$-dimensional real space), for the given data of size $n$. The data set can be clustered into $K$ $(\geq 2)$ hard clusters using our proposed clustering algorithm, for any $p$ $(\geq 1)$ even close to or greater than $n$, utilizing an appropriate distance measure (may not be strictly metric) computed between any two members $M_m$ and $M_{m'}$ as $d(M_m,M_{m'})$. Thus, the proposed cluster analysis is performed throughout under a univariate set-up of the interpoint distances.
	\subsection{An intuitive presentation of the proposed method}
	Firstly, among the members $M_1,\ldots,M_n$, we find the member around which the probability of having observations is the highest, i.e. the area around that member is the densest. Now, the area is specified by a user-given neighborhood, say $h(>0)-$neighborhood around the member, which is formed of the neighboring members within $h$ interpoint distance from the considered member. Then all the members falling in that neighborhood make the first cluster around that member (i.e. the members possessing less than $h$ interpoint distances from that member construct the first cluster around the specified member). Next, we search for the second cluster through the rest of the members in the same way and continue in this manner until no members are left unclustered or we reach the last cluster of one member. If there is any such cluster which is less likely to form a separate group in the data set (say, with cluster size less than $n'$), then we merge it with its nearest cluster of size at least $n'$ ($>0$, a user-defined integer). 
	
	The proximity between any two members is measured in terms of the interpoint distance whose value is always greater than zero for two distinct members. Now, for a positive-valued random variable $X$ following
	a univariate continuous distribution specified by a probability density function
	$f$, the probability of having observations less than $h$ is evaluated as
	\begin{equation}
		P(0< X<h)=\int_{0}^{h}f(x)dx\rightarrow f(h/2) h\hspace{.05in} \text{as}\hspace{.05in} h\downarrow 0,
	\end{equation}
	where $\rightarrow$ means `tends to' and $\downarrow$ indicates `decreases to'.
	
	In our analysis, we apply Equation (1) to the interpoint distances (for which the density function may not be known) so that the probability of having data members in the $h-$neighborhood around a particular member is now obtained in terms of the probability of having members within $h$ interpoint distance from that specified member. Hence, for $f$ unknown, the above probability is estimated as 
	\begin{equation}\label{ep}
		\hat{P}=\hat{f}(h/2) h,
	\end{equation}
	where $\hat{f}$ is an estimate of $f$ based on a random sample $X_1,\ldots,X_n$ of size $n$ drawn from $f$. We propose achieving a suitable kernel density estimator discussed below.
	\subsection{Kernel density estimation}
	Given a real-valued random sample $X_1,\ldots,X_n$ of size $n$ following
	a univariate continuous distribution with probability density function
	$f$. For $f$ unknown, we propose estimation of the density using the classical nonparametric kernel density estimator $\hat{f}$ as follows:
	\begin{equation}\label{kde}
		\hat{f} (x) = \frac{1}{nh} \sum\limits_{i=1}^{n}Ker\bigg(\frac{x-X_i}{h}\bigg),
	\end{equation}
	wherein $h(>0)$ is a smoothing parameter and $Ker(\cdot)$ is a real-valued kernel function satisfying $\int_{-\infty}^{\infty} Ker(x) dx=1$ (Silverman 1986; Wand and Jones 1995; Bandyopadhyay and Modak 2018). Throughout this work, we consider the Gaussian kernel, namely,
	\begin{equation}\label{GK}
		Ker(x)=\frac{1}{\sqrt{2\pi}}\exp\bigg(-\frac{x^2}{2}\bigg).
	\end{equation}
	It is to be noted that we use the same letter `$h$' to denote the neighborhood in Section 2.1 and to represent the smoothing parameter in this Section, because our clustering algorithm is implemented with the same numerical value for both these parameters (discussed in detail under Section 2.3.1).
	\subsection{Clustering algorithm}
	The algorithm is explained as follows.\\
	(a) Consider the $m-$th member $M_m$ from the given data set $D=\{M_1,\ldots,M_n\}$.\\
	(a1) Compute the interpoint distances between the member $M_m$ and all other members in the data set as:\\
	\begin{equation*}
		S_m=\{d(M_m,M_{m'}),m'(\neq m)=1,\ldots,n\},
	\end{equation*}  
	which for given the member $M_m$ (i.e. with conditionally provided $M_m$), is a random sample of size $(n-1)$ whose density function is estimated at the point $h/2$ (see, Equation\eqref{ep}) using the kernel density estimator (Equation 3). It subsequently makes an estimate for the required probability of having observations in the $h-$neighborhood around the member $M_m$, say $\hat P(M_m)$ (see, Equation~\ref{ep}).\\
	(a2) For each data member $M_m$ with $m\in\{1,\ldots,n\}$, we repeat step (a1) and obtain $\hat P(M_m)$ for $m=1,\ldots,n$.\\
	(a3) Find the member for which $\underset{1\leq m\leq n}{\max}\hspace{.05 in}\{\hat P(M_m)\}$ holds. Let the corresponding member be $M_i$, then this member of the data set has the densest $h-$neighborhood around it. \\
	(a4) We select all the members $M_{i'}$ s from the $h-$neighborhood of $M_i$
	to form the first cluster $C_1$, i.e. 
	\begin{equation*} 
		C_1=\{M_{i'}\in D: d(M_i,M_{i'})\in [0,h)\}. 
	\end{equation*}\\
	(b) Remove all the members of $C_1$, say of size $n_1$, and update the data set $D$ with the remaining $n-n_1$ (which is the updated value of $n$ now) unclustered members. Then repeat the steps (a1) through (a4) of the
		algorithm to determine the next cluster $C_2$.\\
	(c) Redo the step (b) in an analogous manner such that we keep obtaining the subsequent clusters $C_3,C_4,$ etc. until no members are left to be clustered, or  there remaining is only one member which is naturally classified in the last cluster.\\
	(d) Thus, we have now $K'$ mutually exclusive and exhaustive clusters $C_1$,\ldots,$C_{K'}$ of sizes $n_1,\ldots,n_{K'}$ respectively. If there is any cluster less likely to form a separate group in the data set, we merge it with the other clusters. For the cluster $C_k$ with $n_k<n'$, where $n'\in\{1,2,...\}$ is user-defined to construct clusters of a minimum size, each member of $C_k$ is reassigned to the closest of all the clusters with sizes $\geq n'$. Clearly for $n'=1$, we can allow singletons as well. Here the closest cluster is that one for which the mean of distances between the member (to be reclustered) and all the members already assigned to the closest cluster at the end of the step (c) is a minimum. For a specified value of $n'$, the merger(s) taking place is/are worthwhile or not is decided by the computed ASW (explained in detail under the following Section 2.3.1). Thus the algorithm terminates with $K$ resulting clusters $C_1$,\ldots,$C_K$ of sizes $n_1,\ldots,n_K(\geq n')$ respectively, where $K\leq K'$.    
	\subsubsection{Choice of the parameters in clustering algorithm}
	The algorithm has two important tuning parameters / hyperparameters $h$ and $n'$ that respectively indicate how dense (perhaps just initially, due to the second phase of our algorithm, as described in step (d) under Section 2.3, where we may achieve $K(< K')$ final clusters from $K'$) and how large clusters we aim to achieve. The values of the parameters are specified by the analyst depending upon the sample at hand. Statistically the values should be varied over a logical range for both the parameters and we eventually choose the value of the pair $(h,n')$ which corresponds to the best clusters found. The quality of the clusters is evaluated in terms of the popular cluster accuracy measure the average silhouette width (abbreviated to ASW, see, Rousseeuw 1987; Kaufman and Rousseeuw 2005), which takes a value from -1 to 1 with a higher value reflecting better cluster analysis. Therefore, clustering with that value of $(h,n')$ should be accepted for which the ASW produces a maximum. For all plausible pairs of values for $(h,n')$, we expect to have consistently robust clustering results for a particular data set with natural clusters. For computational convenience, we apply the algorithm to the interpoint distances after normalizing them so that no distances lie beyond [0,1]. This makes it easier to specify a small value of $h$ in the interval $(0,1)$, because so is desired as per the Equation (1), which also serves for the value of the smoothing parameter from Equation (3). 
	
	This convention of assigning the same value to `$h$’ for the neighborhood and to `$h$’ for the smoothing parameter is being followed, in the context of the nonparametric classical univariate kernel probability density estimator, from the beginning to its recent applications (e.g., see, Silverman 1986; Wand and Jones 1995; Matioli et al. 2018). It comes from the very inherent design of the kernel density estimators which is connected to the naive estimators. The kernel estimator at a point is formed of a sum of individual kernels or bumps, each of which is centered at an observed value and spread over the $h$-neighborhood of that observation (see, Fig.~2.5 in Silverman 1986 and Fig.~1 from Matioli et al. 2018). This is done to keep a balance between the amount of smoothing of data and the formation of the optimal neighborhoods using the observed data, i.e. smoothing at a sampled data member is connected to its neighborhood density relevantly. For a small value of smoothing parameter, the kernel density estimator indicates a multi-modal distribution where the multiple clusters are likely to be formed through our method with the same value of the neighborhood around some data members (see, Figs.~2.5a and 2.6a in Silverman 1986); whereas for a large enough smoothing parameter, only one cluster in terms of a unimodal distribution is indicated (see, Figs.~2.5b and 2.6c from Silverman 1986) where the constituent individual kernels are built with the same larger value for $h$-neighborhood around the given observations, while a cluster of bigger size can be expected from our method. However, our algorithm with an appropriate value of $h$ is capable of revealing the true number of clusters (Fig.~2.6b in Silverman 1986). On the other hand, our clustering algorithm selects any integer-valued $n'(\geq 1)$ facilitating the formation of any sorts of clusters well-defined of reasonable sizes, or with noisy, extreme, outlier-affected, sparely distributed or single observation(s).  
	\subsubsection{Illustration of the algorithm through an example}
\begin{figure}
		\centering
		\includegraphics[width=1\textwidth]{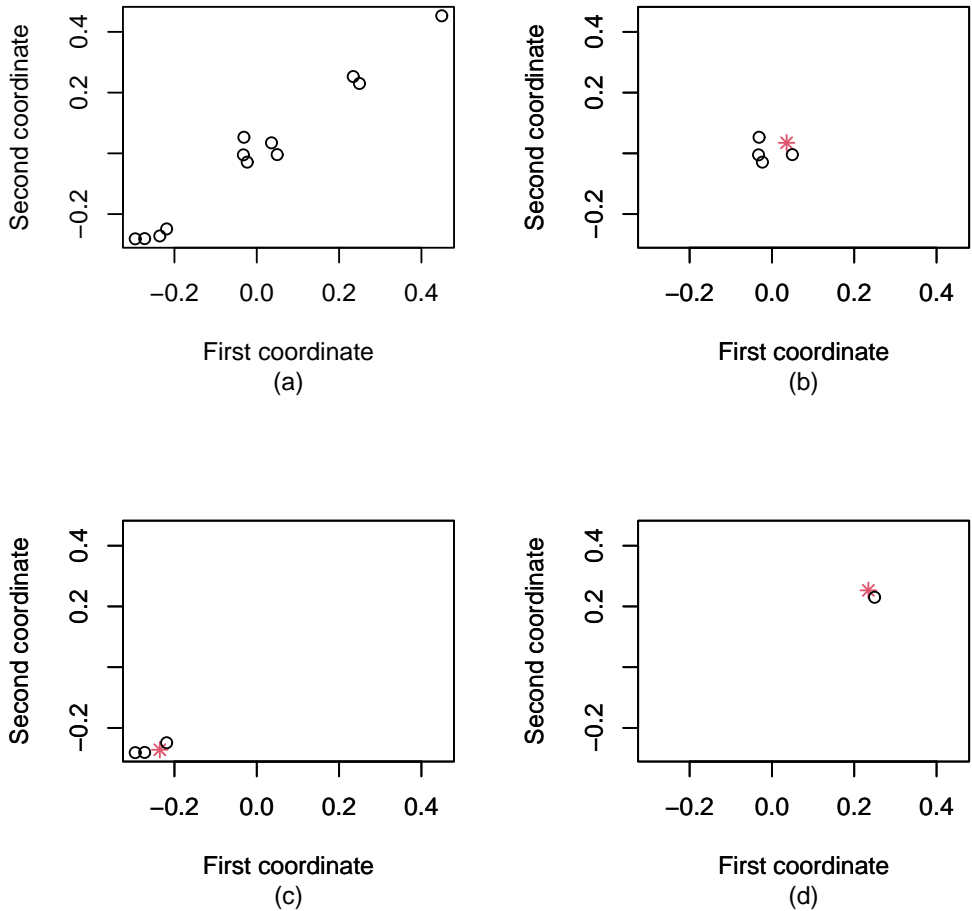}
		\caption{Figures are drawn to illustrate our algorithm step-wise for $h=0.1$. (a) Simulated bivariate data, (b) First cluster, within $h$-neighborhood, around the observation marked with $\textcolor{red}{*}$, (c) Second cluster around $\textcolor{red}{*}$, and (d) Third cluster around $\textcolor{red}{*}$ (without including the last unclustered member of data).}\label{DetailImpl}
	\end{figure}

	We utilize a bivariate simulated data to demonstrate how our cluster algorithm works step by step. Here we specify $h=0.10$. Then, our algorithm from Section 2.3 is implemented numerically as follows (for visual effect, see, Fig.~\ref{DetailImpl}).\\	
	(a: a1--a3) We consider the following data set $D$ with $p=2$ on the first coordinates ($x$) and the second coordinates ($y$) (drawn in Fig.~\ref{DetailImpl}a):\\
		\begin{center}
			\begin{tabular}{c|cccccccccccc}
				\hline				
				$x$&-0.30& 
				-0.22& 
				-0.27& 
				-0.24& 
				-0.03& 
				0.05& 
				-0.03& 
				0.04&  
				-0.02& 
				0.23& 
				0.25& 
				0.45\\
				\hline
				$y$& -0.28& -0.25& -0.28& -0.27&  0.00&  0.00&  0.05&  0.03& -0.03&  0.25&  0.23&  0.45\\
				\hline
			\end{tabular}
		\end{center}
	where $n=12$ and the built-in computation under the present steps achieves:
	\begin{equation*}
		\underset{1\leq m\leq 12}{\max}\hspace{.05 in}\{\hat P(M_m)\}\hspace{.05 in}\text{for}\hspace{.05 in} m=8,\hspace{.05 in} \text{where}\hspace{.05 in} M_8=(0.04, 0.03)'
		.
	\end{equation*}	
	(a4) We obtain $C_1$ with the following members/points which have distances less than 0.01 from $M_8$ (see, Fig.~\ref{DetailImpl}b):
		\begin{center}
			\begin{tabular}{c|ccccc}
				\hline	
				$x$&	-0.03&
				0.05&
				-0.03&
				0.04&
				-0.02\\	\hline
				$y$&0&0&0.05&0.03&-0.03\\
				\hline
			\end{tabular}
		\end{center}		
	(b) Now, the updated data $D$:
		\begin{center}
			\begin{tabular}{c|ccccccc}
				\hline	
				$x$&	-0.30&
				-0.22&
				-0.27&
				-0.24&
				0.23&
				0.25&
				0.45\\	\hline
				$y$&-0.28&-0.25&-0.28&-0.27& 0.25&  0.23&  0.45\\\hline
			\end{tabular}
		\end{center}			
	with updated size $n=7$ and next we find:
	\begin{equation*}
		\underset{1\leq m\leq 7}{\max}\hspace{.05 in}\{\hat P(M_m)\}\hspace{.05 in} \text{corresponding to}\hspace{.05 in} m=4 \hspace{.05 in} \text{with}\hspace{.05 in}  M_4=( -0.24, -0.27)',
	\end{equation*} 
	and consequently, we reach (Fig.~\ref{DetailImpl}c)
	\begin{equation*} 
		C_2=\{ (-0.30, -0.28)',(-0.22, -0.25)',(-0.27, -0.28)',(-0.24, -0.27)'	\}.
	\end{equation*}
	(c) Next, $D=\{(0.23, 0.25)',(0.25, 0.23)',(0.45, 0.45)'
	\}$ with $n=3$,	for which\\
	\begin{equation*}
		\underset{1\leq m\leq 3}{\max}\hspace{.05 in}\{\hat P(M_m)\}\hspace{.05 in} \text{gives}\hspace{.05 in}\hspace{.05 in}m=1 \hspace{.05 in}\text{with}\hspace{.05 in} M_1= (0.23, 0.25)',
	\end{equation*}	
	and we obtain $C_3=\{ (0.23, 0.25)',(0.25, 0.23)'\}$ (Fig.~\ref{DetailImpl}d).
	
	Subsequently, now 
	\begin{equation*}
		D=\{(0.45, 0.45)'\},
	\end{equation*} 
	which is a singleton with $n=1$, therefore, this last member is automatically clustered in $C_4=\{(0.45, 0.45)'\}$.\\
	(d)	For $n'=2$, the merger between the last singleton and its nearest cluster (i.e. $C_3$) gives ASW = 0.75934, whereas, the choice $n'=1$, keeps the singleton as a separate cluster computing ASW =  0.76723. It implies the singleton is an outlier and hence should be clustered as a separate group of 1. Therefore, our algorithm results in an acceptable outcome, for $(h=0.10,n'=2)$, with 4 clusters having cluster memberships of the given data members as: $\{2, 2, 2 ,2 ,1, 1 ,1 ,1 ,1, 3 ,3, 4\}$.
	
	We attach a brief R code to run our algorithm, for its multivariate version, in the appendix.
	
	\section{Numerical experiments}
	To demonstrate the applicability of our method we perform the following data study, wherein, as mentioned earlier, our clustering algorithm can adopt any distance measure (assumed to possess a density function, possibly unknown) depending upon the nature of the given data. For example, as per requirements, we implement the Gower's distance, geodesic and the Euclidean norm (note, if not mentioned otherwise, the distance is Euclidean). For comparison purpose, the same interpoint distance measure is used for a particular data set while carrying out the proposed and its competitive algorithms, namely, the robust $K-$medoids and the popular density-based clustering for arbitrary-shaped clusters DBSCAN, and computing the cluster accuracy measure ASW; however, the classical method $K$-means is only applicable with Euclidean distance from the options considered. We implement $K-$medoids method using the `PAM' algorithm (see, for details, Kaufman and Rousseeuw 2005), $K-$means clustering using the Hartigan--Wong algorithm (Hartigan and Wong 1979) and DBSCAN using a kd-tree (Ester et al. 1996).
	
	The number of clusters $(K)$ present in the set of data under study, whose true value is unknown, is estimated as $\hat{K}$ by our algorithm itself. On the other hand, $K-$means and $K-$medoids algorithms are run for different values of $K$ as a priori with computation of the ASW, and subsequently, that value of $K$ is determined as $\hat{K}$ for which the computed value of the ASW is a maximum; whereas the competitor DBSCAN also evaluates $\hat{K}$ itself, but requires proper selection of the values for its hyperparameters denoted by `$minPts$' and `$\epsilon$'. One reasonable choice would be $minPts=2 \times$ dimension of data, while the value of $\epsilon$ is chosen as that where a knee appears in the curve from a $k$-nearest neighbor ($k$NN) distance plot for $k=minPts$ (see, Ester et al. 1996; Hahsler et al. 2019). The graph displays the $k$NN distance, that is the distance from a member to its $k$-th NN, of all data members sorted from the smallest to the largest, and can be used to help find suitable parameter values for DBSCAN. However, in different situations, this thumb rule may not work and we need to choose the final values using subjective judgment from a range of plausible values.
	
	It is to be noted that in the literature, the clustering results are often reported as what is the estimated value of $K$, but the fact is, in spite of achieving $\hat{K}=K$ correctly by an algorithm the clustering quality may not be good enough to be acceptable. Therefore, we convey the detailed outcome, not only the estimated value for the number of clusters, but also the percentage of misclassification rate, wherever possible. 
	We follow the convention of applying a clustering algorithm to a specific data set, either with known classes or with the assumption of having inherent unknown clusters, for the purpose of evaluation of the efficacy of our clustering method (Ruspini 1970; Hartigan 1975; Kaufman and Rousseeuw 2005; Matioli et al. 2018). Here, we study four simulated data sets, namely S1--S4, and two real-world sets of observations on spatial and biostatistical data, say D1 and D2, respectively. 
	
\begin{figure}
		\centering
		\includegraphics[width=1\textwidth]{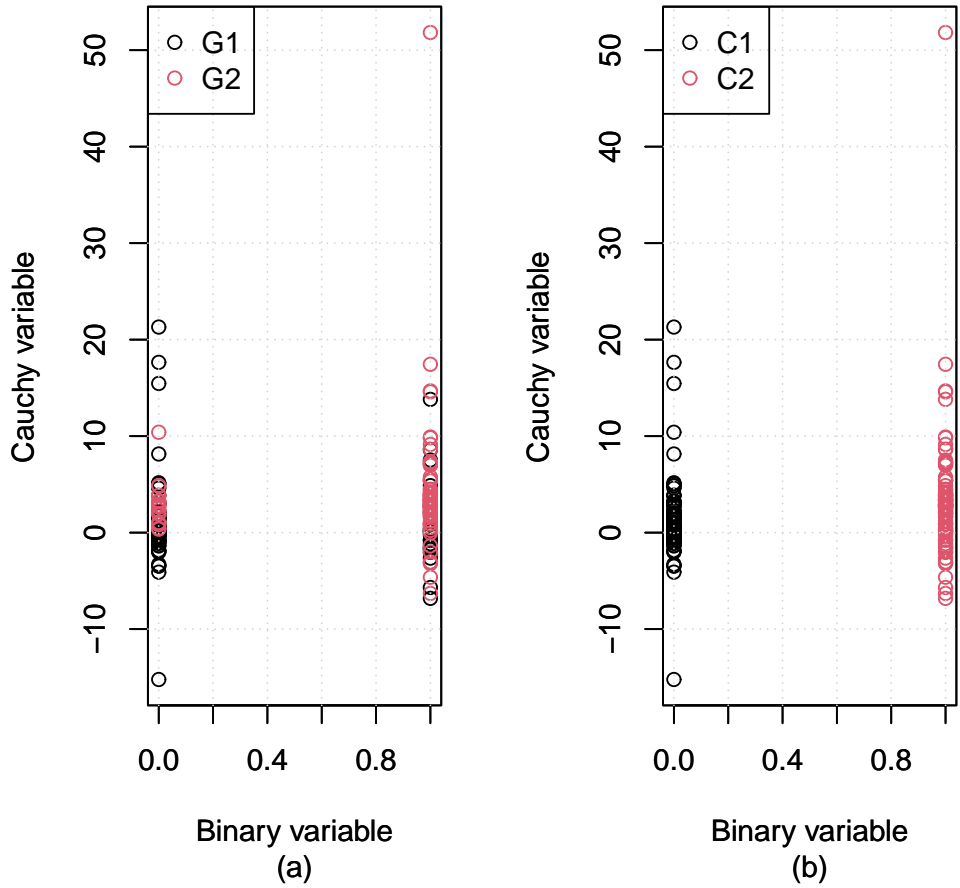}
		\caption{Visual impact of cluster analysis is shown, which is obtained through our proposed clustering algorithm, with $h=0.20$ and $n'=3$, for the simulated bivariate data set (S1), with measurements on mixed scales (namely, binary and continuous), where the first variable (placed along x-axis) is binary taking values 0 or 1, and the second one (drawn along y-axis) is independent Cauchy variable. (a) Plot of sampled data from 2 overlapping groups G1 and G2, and (b) graph of 2 resulted clusters C1 and C2.}\label{MixedScaleBiV}
	\end{figure}

	(S1) Firstly, we generate a data set with two classes having equal sizes 100, involving mixed type of bivariate sample. The first variable is binary, which owns two values 0 and 1 with respective probabilities `$prob$' and $(1-prob)$ for $prob\in (0,1)$, and the second one is a continuous variable, obeying independently a Cauchy distribution with location $\mu$ and scale $\sigma=1$.
	The first class possesses $prob = 0.8$ and $\mu=0$, whereas for the other $prob = 0.2$ and $\mu=3$. The sampled data is shown in Fig.~\ref{MixedScaleBiV}(a), where two inherent groups are quite overlapping and hence it is really challenging to find out the different clusters through cluster analysis.
	
	Our clustering algorithm is applied here with the Gower’s distance which is appropriate for such data measured on arbitrary scales (see, for example, Gower 1971; Kaufman and Rousseeuw 2005). The clustering method, with the hyperparameters $h=0.2$ and $n'=3$, discloses two clusters having an ASW = 0.94090. This high enough value of computed ASW indicates quite good clustering. The clustered data is drawn in Fig.~\ref{MixedScaleBiV}(b), wherein two resulted clusters are prominent. From the graph, it is clear that the outliers coming from the Cauchy distribution are not disturbing the results from our method. This example also demonstrates how our algorithm is applicable to any kind of data by using an appropriate interpoint distance, while most other methods like $K$-means do not have that privilege. 
	
\begin{figure}
		\centering
		\includegraphics[width=1\textwidth]{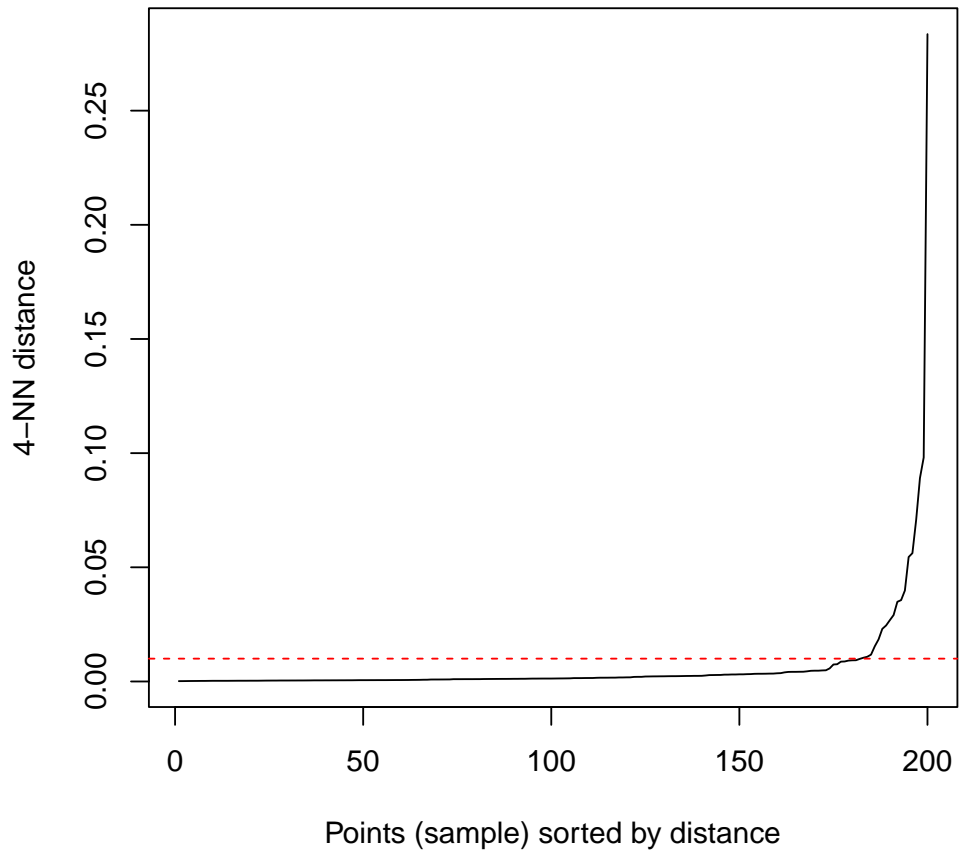}
		\caption{4-nearest neighbor distance plot indicates that $\epsilon=0.01$ can be considered, as the red line approximately represents the knee in the curve, for the DBSCAN algorithm with $minPts=4$.}\label{DGS1}
	\end{figure}

	As a result, $K$-means through Euclidean norm reveals the two groups with much lesser accuracy, indicated by ASW = 0.79527; whereas the $K$-medoids with Gower’s distance, known to be robust under outliers, comes out to be competitive with our method, having equal efficacy (see, Table \ref{t:newt1}, highest values for ASWs are indicated in bold, for ease of understanding). On the other hand, DBSCAN (see, Fig.~\ref{DGS1}), in association with Gower’s distance, is surpassed by our algorithm, producing ASW = 0.68255 with mainly two clusters of sizes 94 and 91, and a third group of 4 members with the remaining 11 data members marked as noise. This is the problem of DBSCAN algorithm, when the analyst needs each member of the data to be classified into some cluster, the resulting noises may involve subjectivity for assignment of their respective cluster-memberships.
	
\begin{figure}
		\centering
		\includegraphics[width=1\textwidth]{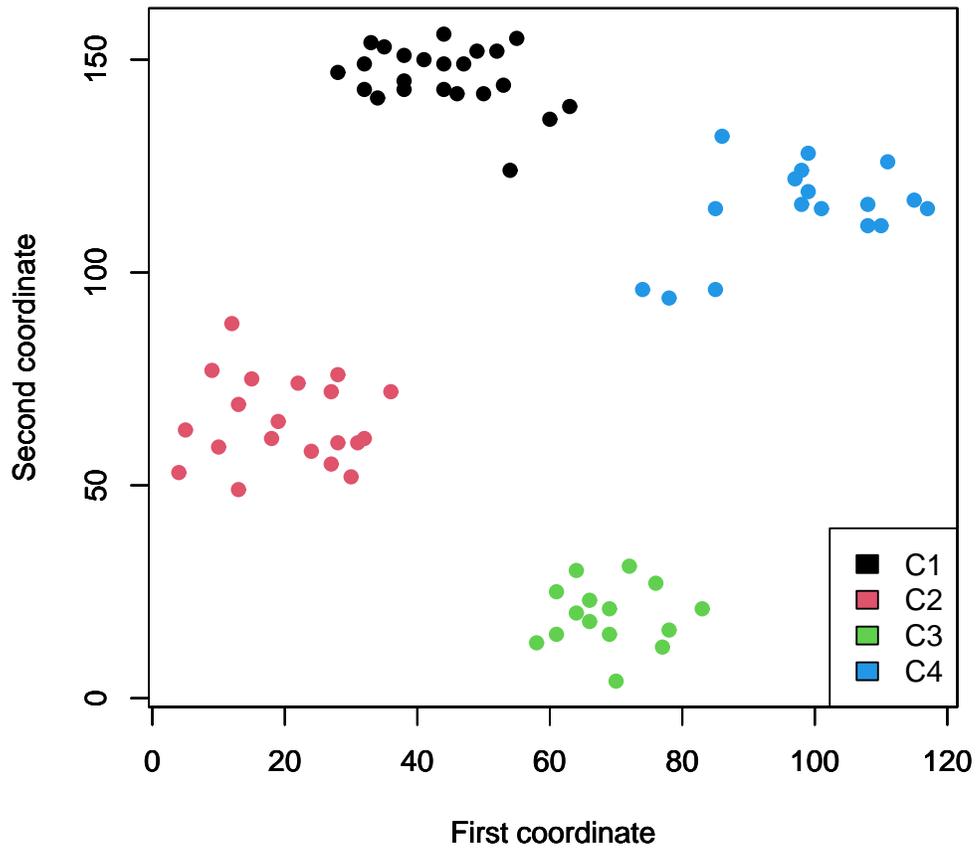}
		\caption{Existing four classes in the Ruspini data set (S2), identified by our clustering algorithm, as clusters C1--C4, for $h=0.10$ and $n'=4,5$, with 100\% accuracy.}\label{Ruspini}
	\end{figure}

	(S2) Secondly, we use `Ruspini' data set (Ruspini 1970), a benchmark for illustrating clustering techniques. Originally, it was used by Dr. Enrique H. Ruspini, in the context of evaluation of the performances for different fuzzy clustering methods. Now this data set is popularly applied for both fuzzy and hard clustering algorithms. It is a simulated set of observations with four classes which consist of the coordinates of 75 points in two-dimensional space as shown in Fig.~\ref{Ruspini}. 
	
	We perform the cluster analysis using our algorithm with different values for the parameters as $(h,n')\in\{0.10, 0.15, 0.20, 0.25, 0.30\}\times\{ 3, 4, 5\}$ to identify the inherent clustering structure. The results are reported in Table~\ref{t:t1}, wherein the highest value of the ASW is attained for $h=0.10$ and $n'=4$, $5$ (optimal values of ASW are highlighted in bold), which give rise to the correctly identified four clusters existing in the Ruspini data set with 100\% accuracy. 
	
	$K-$medoids method also discovers the same with $\hat{K}=4$ in terms of the ASW, whereas
		$K-$means clustering fails with indication of three clusters (i.e. $\hat{K}=3$), and so does partly DBSCAN method ($\epsilon=17,minPts = 4$) generating fours clusters but three noise points. While assuming that we do not know the original groups of this data set, i.e. under unsupervised classification, the computed ASW values clearly lead us to which algorithm's answer is to be accepted, e.g. (Algorithm, ASW) = (Our method, 0.73766), ($K$-medoids, 0.73766), ($K$-means, 0.64139
		), (DBSCAN,  0.71348). It exhibits that our method and $K$-medoids algorithm are producing the best results, and now when we compare the clustering results with the actually known class memberships of the data, we find out only these two methods expose the real four clusters with 100\% accuracy. 
	
	(S3) Next, we simulate a data set from a six-variate normal population with the following complex multivariate structure established by a $t-$copula. The copula is defined by a six-variate $t-$distribution having $2$ degrees of freedom and the correlation matrix with $0.15$ as the off--diagonal entries (Nelsen 2006; Modak and Bandyopadhyay 2019). For $T$ representing the distribution function of the above-mentioned multivariate $t-$distribution and $T_i$ producing the corresponding marginal distribution function for the $i-$th variable with the inverse function $T_i^{-1}$, the $t-$copula is given by
	\begin{equation*}
		C(u_1, . . . , u_{6}) = T\{T_1^{-1} (u_1), . . . , T_{6}^{-1}(u_{6})\}, 0 < u_i < 1, i = 1, . . . ,6.
	\end{equation*}
	We create three different groups of sizes 20, 15 and 10 with all six variables, under the above-described multivariate structure, following the marginal distributions $N(0,1)$, $N(-3,1)$ and $N(3,1)$ respectively.
	
\begin{figure}
		\centering
		\includegraphics[width=1\textwidth]{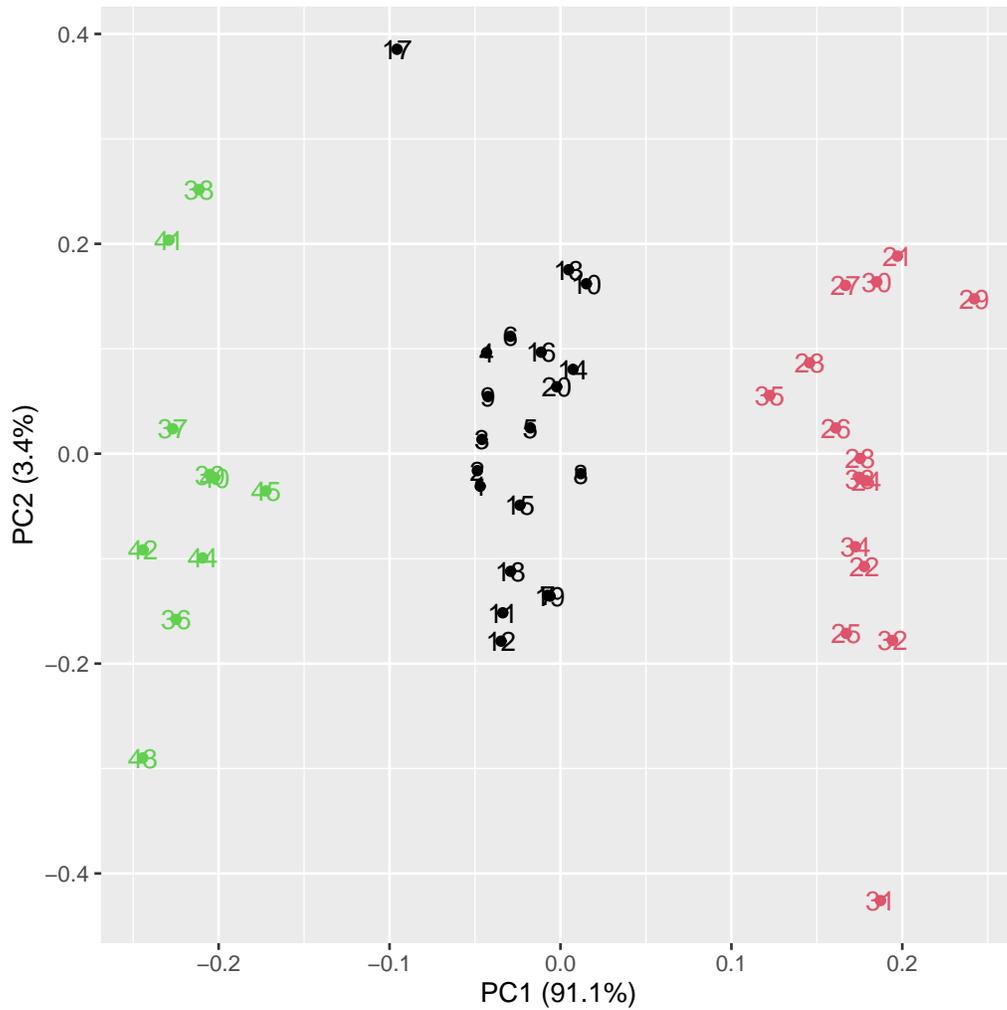}
		\caption{2D plot of multivariate normal data (S3) with respect to the first principal component
			(PC1, with 91.1\% variation) versus the second principal component
			(PC2, with 3.4\% variation), where cluster 1
			(black), cluster 2 (red) and cluster 3
			(green) of respective sizes $(20,15,10)$, as resulted by our algorithm  $(h=0.15,n'=3)$ with 100\% accuracy, are plotted with the digits printed as indices of the sampled observations.}\label{PCA_MulNormal}
	\end{figure}

	Table~\ref{t:t2} shows the clustering output from our algorithm in detail. Three clusters are indicated with the higher values of the ASW, whereas the highest values of the ASW correspond to better clustering with 100\% correct classification rate. For example, the proposed algorithm with $(h,n')=(0.30,3)$ produces the ASW = 0.61226, which is less than the ASW = 0.64155 resulted in our method for $(h,n')=(0.15,3)$. These two analyses respectively generate 97.78\% and 100\% correct classification rates. For the latter, a visual interpretation is given in Fig.~\ref{PCA_MulNormal}, where we can prominently see the three independent clusters for this multivariate data set in terms of the first two orthogonal principal components, achieved through the popular dimension reduction technique, namely the classical linear principal component analysis (Sch\"{o}lkopf and Smola 2002; Modak et al. 2018). Thus it is clearly proven that, for our algorithm, the values of $h$ and $n'$ can be chosen by the resulted values of the ASW, whose larger value suggests better clustering. The detailed output (see, Table~\ref{t:t2}) precisely uncover the three distinct clusters explored by our algorithm, consistently for most of the considered pairs of values for $(h,n')$.
	
	The competitors under our consideration $K$-means and $K$-medoids methods are constructed in a way that they are good enough in recognizing normal clusters. As a result, both of them robustly satisfy our outcome, whereas DBSCAN $(\epsilon = 2.95, minPts = 5)$ gives three clusters of sizes 19, 15, 10 and 1 noise point for ASW = 0.54415, which decreases its performance in decision making through clustering the present data set.
	
	Now, as this example under the considered multivariate structure resembles a general real-life situation, we perform an extensive study on the computation time in this scenario. Using the function `SM\_algorithm' (see, appendix \ref{algo}), constructed in the global statistical `R' programming language, we run our clustering algorithm with hyperparameters $(h,n')=(0.15,3)$ under the above set-up. It has the CPU time (in seconds for difference between the start and the end of the program) reported for different $p$ and size of each of the three groups $n_0$ (say, i.e. total sample size $=3n_0$) in Table~\ref{t:time}. The resource details are listed as follows: operating system - windows 10, processor - intel core i3, laptop - 64 bit, RAM - 4gb, R version 4.1.0. It is shown to be fast enough, specially, for big data, and high-dimension with low sample size situations (known as HDLSS, see Chen and Qin 2010; Marrozi 2015).
	
\begin{figure}
		\centering
		\includegraphics[width=1\textwidth]{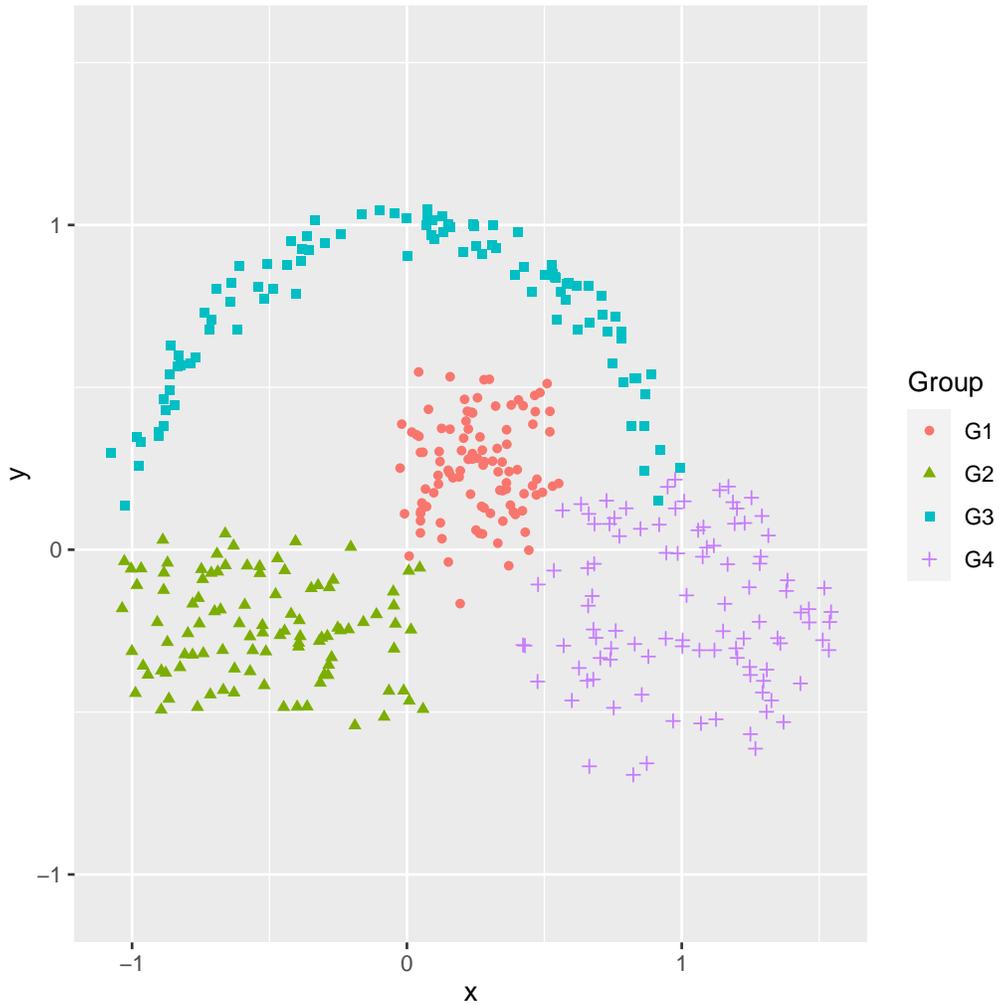}
		\caption{Simulated data set (S4) with four groups $G1-G4$, each with random sample of size 100, independently drawn from a square, rectangle, half-circle and circle shaped distributions, respectively, in a bivariate space where each variable is contaminated with Gaussian noise possessing mean 0 and standard deviation 0.05.}\label{NoisyDiffShape}
	\end{figure}

	(S4) The last simulation study manifests how our proposed method can retrieve challenging clusters which are extremely close to each other and having completely different shapes with noisy observations. For this purpose, we use the second synthetic data set from Modak (2022a), wherein a four-group situation, each with size 100, is considered in a bivariate set-up (see, our Fig.~\ref{NoisyDiffShape}, that is a recreation of Fig.~2 from Modak 2022a).
	
	This big data set of size 400 is clustered with our fast and easy-to-implement method, where the pair $(h=0.25,n'=27)$ gives 4 clusters of sizes  161, 101, 82 and 56
		with ASW = 0.38290. As a competitor, we expect the popular density-based DBSCAN algorithm to be competitive enough in the present situation (Ester et al. 1996; Campello et al. 2013; Modak 2022a), due to its efficiency to expose arbitrary-shaped clusters, unlike the other used rivals $K$-means or $K$-medoids, and its robustness in the presence of noises. However, DBSCAN method, with values of its hyperparameters $\epsilon=0.165$ and $minPts = 10$, extracts 4 clusters  of 275, 53, 39 and 19 with 14 noise members for an ASW 0.02478. As far as the other competitors are concerned, the classical $K$-means actually becomes competitive for the given set of data with estimating $K$ correctly as 4 for a value of ASW = 0.48457, generating groups having 155, 102, 100 and 43 members. However,
		$K$-medoids turns out to be unsuccessful in the present situation, with indication of 5 well-sized clusters of 110, 99, 98, 53 and 40.
	
	(D1) Data analysis using spatial effect reveals a lot about the sample given (Modak et al. 2017; Matioli et al. 2018). Our fifth data study is carried out on the geographical
	coordinates in terms of longitude and latitude corresponding to the first 60 lightning happened in the year 2011 in the South and Southeast regions of the country Brazil. The bivariate spatial data set is collected from Table 6 of Matioli et al. (2018).
	
\begin{figure}
		\centering
		\includegraphics[width=1\textwidth]{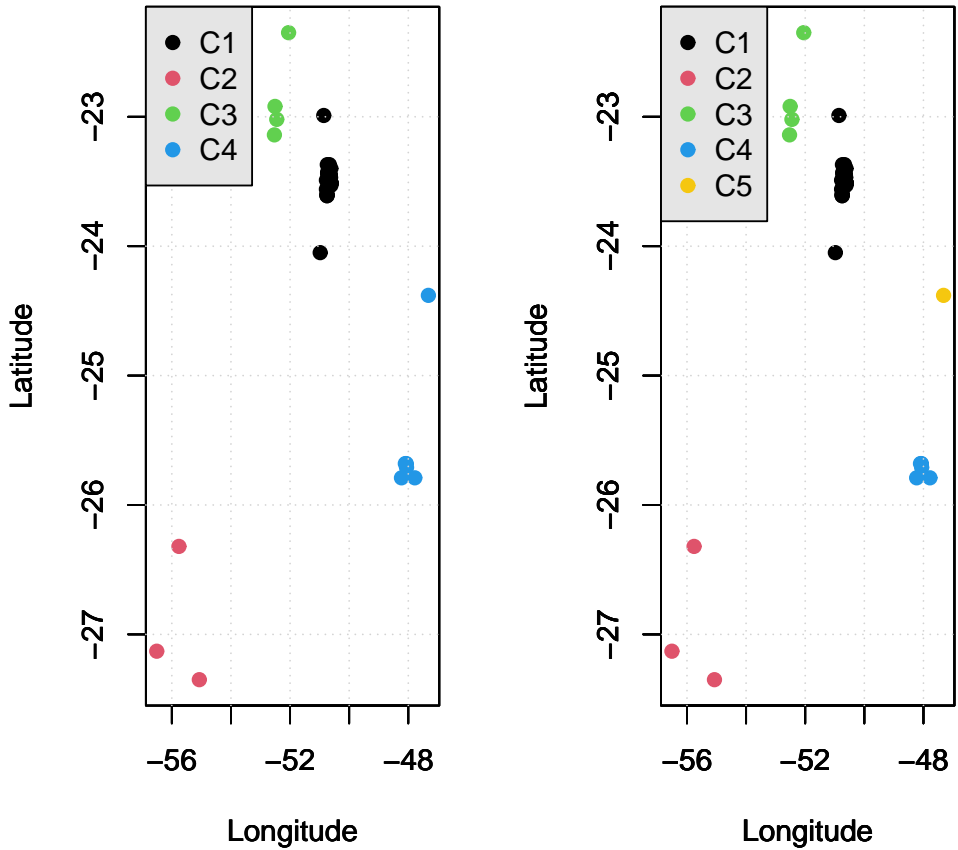}
		\caption{(a) 2D scatter plot of the clustered spatial data (D1) for 60 places as determined by our algorithm with $(h=0.15,n'=3)$, and (b) 2D scatter plot of the clustered spatial data (D1) for 60 places as reached by both our algorithm with $(h=0.15,n'=1)$ and the ClusterKDE algorithm with $(\alpha_1=\alpha_2=5,h_1=h_2)$.}\label{spatial}
	\end{figure}

	Our Table~\ref{t:t3} shows the proposed algorithm with a highest value of the ASW = 0.89635 (marked in bold), for $h=0.15$ and $n'=3$, results in $\hat{K}=4$. It exposes four clusters of lightning which are spatially close (see, Fig.~\ref{spatial}a). This result is robustly confirmed by both the $K-$means and $K-$medoids clustering methods, which generate the highest values of the ASW for given $K=4$, i.e. they also give rise to $\hat{K}$ as four (see, Table~\ref{t:t4} wherein the optimal values of ASW are shown
	in bold). However, DBSCAN $(\epsilon = 0.22, minPts = 3)$, with ASW = 0.75373, performs worse, making three clusters with 45, 4 and 3 data members, and treating the rest of 8 members as noise.
	
	Interestingly, we compare our results with those reported in Matioli et al. (2018), from where we retrieve the present data. Their implemented kernel-based clustering algorithm, namely ClusterKDE (bidimensional version, see, Matioli et al. 2018) wherein $K$ is not needed as a priori, with hyperparameters $\alpha_1=\alpha_2=5$ and $h_1=h_2$, makes partitions that coincide with those from our algorithm for $h=0.15$ and $n'=1$. However, ClusterKDE gives 5 clusters corresponding to a lesser ASW value 0.88726 (see, Fig.~\ref{spatial}b), which is outperformed by our optimal results with 4 clusters, i.e. by our algorithm for $h=0.15$ and $n'=3$ (see, Table~\ref{t:t3}). It is to be noted that we do not use this `ClusterKDE' algorithm as a competitor, in general, because its application is restricted to a maximum of 2-dimensional space, and most importantly, for different values of its hyperparameters, the algorithm does not always converge.
	
	As the data concern spatial observations, we also study our method in association with a spatial distance measure, namely geodesic (see, for reference, Karney 2013), which gives a highly accurate estimate of the shortest distance between any 2 points on an ellipsoid. This interpoint distance leads us consistently to the same results of 4 clusters through our method, for the above-mentioned hyperparameters $(h=0.15,n'=3)$, with ASW = 0.89314. These results are also verified by the partitioning-based clustering algorithm $K-$medoids, while the density-based method DBSCAN $(\epsilon =25000, minPts = 3)$ is outrun having an ASW = 0.75431, with the exact partitions as in the case of Euclidean norm. Thus, for these algorithms, we can confirm the clustering results with respect to different distances, whereas $K$-means is not applicable along with this interpoint distance (note: the observed difference in ASW values, for the same resulting clusters, are due to distinct distance formulas).
	
	(D2) The last application involves a high-dimensional biostatistical data set (i.e. `Alon' data, see, for details, Alon et al. 1999) with sample size 62 and 2000 variables, wherein two inherent groups are known to be present. Here individual class level is available which helps evaluate the performance of our clustering algorithm precisely. One group consists of 40 patients diagnosed with colon cancer and the other has 22 healthy patients. 
	
	We cluster these 62 patients, based on their measured 2000 genes as study variables, where our novel approach, for $h=0.3$ and $n'=3$, unveils two clusters (i.e. $\hat{K}=2$) with 66.129\% correct classification rate. On the other hand, $K-$means and $K-$medoids algorithms, for two number of clusters provided as a known priori (i.e. $K=2$ is specified), give only 62.903\% and 46.774\% correct classification rates, respectively. However, if $K$ is unspecified, like generally happens in the cluster analyses, and to be selected in the usual way, then ASW values imply that $K$-means wrongly estimates $K$ as three, although $K$-medoids does correctly. Here the dimension is much higher than the sample size, therefore, we try for different  values of tuning parameters in DBSCAN rather than following the thumb rule, where DBSCAN algorithm turns out to be not capable enough to explore the inherent clustering pattern. It mainly finds one cluster and the other as a group of noises, with a poor classification rate with respect to the original known classes, for different choices of its hyperparameters. For example, with $minPts=5$, a knee is found at $\epsilon=16350$, that results in only one cluster of 41 members and 21 noisy ones with a mere $38.710\%$ successful classification.
	\section{Conclusion}
	In this paper, we propose a new interpoint distance-based nonparametric
	clustering algorithm which can classify sets of data, measured on arbitrary scales in any dimensional space, using the user-defined distance measure. The chosen interpoint distance is assumed to possess a density function (not necessarily known), which is estimated by the classical nonparametric univariate kernel density estimator and then is used to find the densest neighborhood around a data member to construct a cluster. Our clustering algorithm is simple in its formation and easy to apply results in well-defined clusters. Our suggested approach objectively selects the initial cluster representative and always converges irrespective of it and of the chosen values for its tuning parameters. The method determines the number of clusters existing in the data by itself during the procedure. Wide applicability,
	high-dimensional use, efficient performance even with outliers, noisy observations or arbitrary-shaped clusters, and supremacy relative to the well-known competitors, confirmed by our extensive data study, manifest the novel method as a strong and useful algorithm for clustering purpose. Being an interpoint distance-based approach, of course, the computation time increases with an increase in the data size. Also, some automatic (possibly empirical) technique, intrinsic to the design of the proposed algorithm, would be interesting to be suggested for choosing the values of the hyperparameters associated with our new algorithm. In this kernel-based approach, any kernel other than the presently implemented Gaussian one can be used, and their performance could be fascinating to study while incorporating some other clustering accuracy measures in the algorithm as well than the ASW.
	\section*{Acknowledgments}
	The author would like to express her sincere gratitude to the Editor-in-chief to encourage the work. Author greatly acknowledges an anonymous associate editor for meticulous judgment, proficient advice, considerate appreciation of the significance regarding the present work and for giving the opportunity for its revision. The author feels short of words to thank enough the three esteemed (anonymous) reviewers to read the manuscript to its intrinsic details and kindly provide their expert feedback that helped to draw the author's attention to the potential improvements and to revise the manuscript to a substantial degree which increased its exposition.
	\section*{Disclosure statement}
	No potential conflict of interest was reported by the author.
	\clearpage
	
	\begin{table}
		\caption{Clustering results obtained through $K-$means and $K-$medoids algorithms at different values of the number of clusters ($K$), as a priori specified, for data set (S1)} 
		
		\begin{center}
			\begin{tabular}{ccc}
				\hline\\			
				$K$&ASW&ASW\\
				& ($K-$means)&($K-$medoids)\\[1ex]
				\hline\\	
				
				2&\textbf{0.79527}&\textbf{0.94090}\\
				3&0.52358& 0.78895\\
				4&0.51760& 0.57644\\
				5& 0.40993& 0.53655\\
				6& 0.47750& 0.56084\\
				\hline
			\end{tabular}
		\end{center}
		\label{t:newt1}
	\end{table}
	\clearpage
	\begin{table}
		\caption{Clustering results obtained through our algorithm for Ruspini data set (S2)} 
		\begin{center}
			\begin{tabular}{ccc}
				\hline\\			
				($h,n'$)&ASW&$\hat{K}$\\[1ex]
				\hline\\		     			
				
				(0.10,    3)& 0.57708&    8\\[1ex]
				(0.10,    4)& \textbf{0.73766}&    4\\[1ex]
				(0.10,    5)& \textbf{0.73766}&    4\\[1ex]
				(0.15,    3)& 0.47460&    6\\[1ex]
				(0.15,    4)& 0.47460&    6\\[1ex]
				(0.15,    5)& 0.47460&    6\\[1ex]
				(0.20,    3)& 0.55417&    5\\[1ex]
				(0.20,    4)& 0.55417&    5\\[1ex]
				(0.20,    5)& 0.55417&    5\\[1ex]
				(0.25,    3)& 0.67136&    4\\[1ex]
				(0.25,    4)& 0.67136&    4\\[1ex]
				(0.25,    5)& 0.67136&    4\\[1ex]
				(0.30,    3)& 0.66779&    4\\[1ex]
				(0.30,    4)& 0.66779&    4\\[1ex]
				(0.30,    5)& 0.66779&    4\\[1ex]
				\hline
			\end{tabular}
		\end{center}
		\label{t:t1}
	\end{table}
	
	\clearpage
	\begin{table}
		\caption{Clustering results obtained through our algorithm for  multivariate normal data set (S3)} 
		\begin{center}
			\begin{tabular}{ccccc}
				\hline\\			
				$(h,n')$&ASW&$\hat{K}$& Cluster&Correct \\
				&   &         &  sizes&classification \\
				&   &         &       &  rate \\[1ex]
				\hline\\

				(0.10,    3)& 0.39423&    4&(13, 15,  7, 10)&84.44\%\\[1ex]
				(0.10,    4)& 0.39423&    4&''&''\\[1ex]
				(0.10,    5)& 0.39423&    4&''&''\\[1ex]
				(0.15,    3)& 0.64155&    3&(20, 15, 10)&100.00\%\\[1ex]
				(0.15,    4)& 0.64155&    3&''&''\\[1ex]
				(0.15,    5)& 0.64155&    3&''&''\\[1ex]
				(0.20,    3)& 0.64155&    3&''&''\\[1ex]
				(0.20,    4)& 0.64155&    3&''&''\\[1ex]
				(0.20,    5)& 0.64155&    3&''&''\\[1ex]
				(0.25,    3)& 0.64155&    3&''&''\\[1ex]
				(0.25,    4)& 0.64155&    3&''&''\\[1ex]
				(0.25,    5)& 0.64155&    3&''&''\\[1ex]
				(0.30,    3)& 0.61226&    3&(21, 14, 10)&97.78\%\\[1ex]
				(0.30,    4)& 0.61226&    3&''&''\\[1ex]
				(0.30,    5)& 0.61226&    3&''&''\\[1ex]
				\hline
			\end{tabular}
		\end{center}
		\label{t:t2}
	\end{table}
	\clearpage
	\begin{table}
		\caption{Run time of our algorithm on the data described under (S3)} 
		\begin{center}
			\begin{tabular}{|ccc|}
				\hline
				
				Time&Data&Size of each\\
				(difference&dimension&sampled group\\
				in seconds)  	&$(p)$&$(n_0$)\\\hline			
				$<1$ &2&50\\
				1&2&100\\
				3&2&250\\
				12&2&500\\
				$<1$ &5&50\\
				1&5&100\\
				3&5&250\\
				12&5&500\\
				$<1$ &10&50\\
				1&10&100\\
				3&10&250\\
				14&10&500\\
				$<1$ &50&50\\
				1&50&100\\
				5&50&250\\
				21&50&500\\
				$<1$&500&50\\
				24&500&250\\
				1&1000&50\\
				8&1000&100\\
				55&1000&250\\
				3&2000&50\\
				16&2000&100\\
				118&2000&250\\
				\hline
			\end{tabular}
		\end{center}
		\label{t:time}
	\end{table}
	\clearpage
	\begin{table}
		\caption{Clustering results obtained through our algorithm for bivariate spatial data set (D1)} 
		\begin{center}
			\begin{tabular}{ccc}
				\hline\\			
				($h,n'$)&ASW&$\hat{K}$\\[1ex]
				\hline\\		     			
				
				(0.10,    3)& 0.82943&    3\\[1ex]
				(0.10,    4)& 0.82943&    3\\[1ex]
				(0.10,    5)& 0.78539&    2\\[1ex]
				(0.15,    3)& \textbf{0.89635}&    4\\[1ex]
				(0.15,    4)& 0.82943&    3\\[1ex]
				(0.15,    5)& 0.78539&    2\\[1ex]
				(0.20,    3)& 0.88863&    3\\[1ex]
				(0.20,    4)& 0.78539&    2\\[1ex]
				(0.20,    5)& 0.78539&    2\\[1ex]
				(0.25,    3)& 0.78539&    2\\[1ex]
				(0.25,    4)& 0.78539&    2\\[1ex]
				(0.25,    5)& 0.78539&    2\\[1ex]
				(0.30,    3)& 0.78539&    2\\[1ex]
				(0.30,    4)& 0.78539&    2\\[1ex]
				(0.30,    5)& 0.78539&    2\\[1ex]
				
				\hline
			\end{tabular}
		\end{center}
		\label{t:t3}
	\end{table}
	
	\clearpage
	\begin{table}
		\caption{Clustering results obtained through $K-$means and $K-$medoids algorithms at different values of the number of clusters ($K$), as a priori specified, for bivariate spatial data set (D1)} 
		\begin{center}
			\begin{tabular}{ccc}
				\hline\\			
				$K$&ASW&ASW\\
				& ($K-$means)&($K-$medoids)\\[1ex]
				\hline\\	
				
				2&0.77510&0.78539\\[1ex]
				3&0.88863&0.88863\\[1ex]
				4&\textbf{0.89635}&\textbf{0.89635}\\[1ex]
				5&0.78059&0.88727\\[1ex]
				6&0.66589&0.85505\\[1ex]

				\hline
			\end{tabular}
		\end{center}
		\label{t:t4}
	\end{table}
	\clearpage
	
	\clearpage
	\clearpage
	\clearpage
	\clearpage
	\clearpage
	\clearpage
	\clearpage
	\appendix
	\section{Constructed R function `SM\_algorithm' to run our proposed algorithm under multivariate set-up}\label{algo}
	\verb!##! Attach the following library to compute ASW\\
	\verb!library(cluster) !\\   
	\hfill \break
	\verb!##! User-defined kernel (here Gaussian kernel as in Eq. (4))\\
	\verb!ker<-function(x){!\\
		\verb!	return(exp(-(x^2)/2)/sqrt(2*pi))}!\\
	\hfill \break
	\verb!##! User-defined kernel density estimator as in Eq. (3)\\
	\verb!f <- function (x,y,h)!\\
	\verb!{!\\
		\verb!	estpdf<-sum(ker((x-y)/h))/(length(y)*h)!\\
		\verb!	return(estpdf)!\\
		\verb!}!\\
	\hfill \break
	\verb!##! Working data set is stored in a matrix named `data' with number of rows $n$ and number of columns $p$\\
	\verb!p<-ncol(data)!\\
	\hfill \break
	\verb!##! Compute the interpoint distance matrix (here Euclidean norm is used)\\
	\verb!distMatrix<-dist(data);M<-as.matrix(distMatrix)!\\
	\verb!##! Normalize all the interpoint distances\\
	\verb!M<-M/max(M)!\\
	\hfill \break
	\verb!##! Our algorithm with hyperparameters $h$ and $n'$ is built in the function `SM\_algorithm'\\
	
	\verb!SM_algorithm<-function(h,n_dash){ !\\
		\verb!ASW<-NULL;cl<-rep(NA,nrow(data));data1<-data;M1<-M;index<-0;!\\
		\verb!lowden<-NULL!\\
		\hfill \break
		\verb!##! First round of algorithm stops when either all members are clustered or only one member is left unclustered, as in step (c) under Section 2.3\\
		\verb!while(length(data1)>p)!\\ 
		\verb!{!\\
			\verb!	index<-index+1!\\
			\hfill \break
			\verb!##! Find the index of member with the densest $h$-neighborhood\\
			\verb!	mm<-which.max(apply(M1,1,function(x)f(h/2,x,h=h)))!\\ 
			\verb!##! Isolate all members within the above-said neighborhood as a cluster
			\verb!exclude<-which(M1[mm,]<h)!\\                           
			\verb!for(j in 1:length(exclude))!\\
			\verb!cl[which(apply(data,1,function(x)!\\
			\verb!sum((data1[exclude[j],]-x)^2)==0))]<-index!\\
			\hfill \break
			\verb!##! Update the data set\\
			\verb!data1<-data1[-exclude,] !\\     
			\verb!##! Update the distance matrix\\                  
			\verb!	M1<-as.matrix(dist(data1));M1<-M1/max(M1)!\\ 
			\hfill \break    
			\verb!##! While loop ends with either no member or 1 \\
			\verb!	}!\\  
		\verb!##! If 1 member left unclassified is assigned to the last cluster \\ 
		\verb!cl[is.na(cl)]<-index+1!\\
		\hfill \break
		\verb!##! $K'$ clusters are obtained till now, where \verb!cl! prints the cluster memberships\\
		\hfill \break
		\verb!##! Second round of the algorithm starts to check if merger is needed to get final $K$ clusters for a given value of $n'$, as described in step (d) under Section 2.3\\
		\verb!clus<-sort(unique(cl));K<-length(clus)!\\
		\verb!##! Compute cluster-wise indices\\
		\verb!clusindex<-lapply(clus,function(j)which(cl==clus[j]))!\\  
		\hfill \break
		\verb!##! Get high probable clusters\\
		\verb!highden<-which(as.vector(lapply(clus,function(j)!\\
		\verb!(length(unlist(clusindex[j]))>(n_dash-1))))==TRUE)!\\
		\hfill \break
		\verb!##! Find low probable clusters\\
		\verb!for(j in 1:K)!\\
		\verb!if(length(unlist(clusindex[j]))<n_dash)!\\ \verb!lowden<-c(lowden,unlist(clusindex[j]))!\\
		\hfill \break
		\verb!##! Check for merger of low density clusters with high density ones\\
		\verb!	if (length(lowden)>0) {!\\
			\verb!	for(l in 1:length(lowden))!\\
			\verb!	cl[lowden[l]]<-which.min(lapply(highden,function(j)!\\
			\verb!	mean(M[lowden[l],unlist(clusindex[j])])))} else {!\\
			\verb!	print("No rare clusters")!\\
			\verb!	}!\\
		\hfill \break
		\verb!##! Compute the ASW for the clustered data\\
		\verb!ASW<-mean(silhouette(cl,distMatrix)[,3])!\\ 
		\hfill \break   
		\verb!##! Give outcome with computed ASW and the final cluster memberships\\
		\verb!	print(list(ASW,cl))} !\\   
	\hfill \break                 
	\verb!##! Run our clustering algorithm on a given data with $h=h0$ and $n'=n0$\\
	\verb!SM_algorithm(h=h0,n_dash=n0)!\\    
	

\clearpage
	\begin{thebibliography}{}
		\bibitem{}
		Alon, U., Barkai, N., Notterman, D. A., Gish, K., Ybarra, S., Mack, D. and Levine, A. J. (1999). \textsl{Broad patterns of gene expression revealed by clustering analysis of tumor and normal colon tissues probed by oligonucleotide arrays}. Proceedings National Academy of Sciences, USA. \textbf{96}, 6745--6750.	
		\bibitem{}
		Arias-Castro, E., Mason, D. and Pelletier, B. (2016). \textsl{On the estimation of the gradient lines of a density and the consistency of the mean-shift algorithm.} Journal of Machine Learning Research. \textbf{17}, 1487--1514.
		\bibitem{}
		Bandyopadhyay, U. and Modak, S. (2018). \textsl{Bivariate density estimation
			using normal-gamma kernel with application to astronomy}. Journal of Applied Probability and Statistics. \textbf{13}, 23--39.	
		\bibitem{}
		Bezdek, J. C. (1981). \textsl{Pattern Recognition with Fuzzy Objective Function Algorithms}. Plenum Press, New York.
		\bibitem{}
		Campello, R. J. G. B., Moulavi, D., Sander, J. (2013). \textsl{Density-Based Clustering Based on Hierarchical Density Estimates}. Proceedings of the 17th Pacific-Asia Conference on Knowledge Discovery in Databases (PAKDD 2013). Lecture Notes in Computer Science. Springer, Berlin, Heidelberg. \textbf{7819}, 160--172.
		\bibitem{}
		Chen, S. X. and  Qin, Y.-L. (2010). \textsl{A two-sample test for high-dimensional data with applications to gene-set testing}. The Annals of Statistics. \textbf{38}, 808--835.
		\bibitem{}
		Cheng, D., Zhu,  Q., Huang, J., Wu, Q. and Yang, L. (2021).  \textsl{Clustering with Local Density Peaks-Based Minimum Spanning Tree}. IEEE Transactions on Knowledge and Data Engineering. \textbf{33}, 374--387.
		\bibitem{}
		Dunn, J. C. (1974). \textsl{Well-separated clusters and optimal fuzzy partitions.}  Journal of Cybernetics. \textbf{4}, 95--104.
		\bibitem{}
		Ester, M., Kriegel, H.-P., Sander, J. \& Xu, X. (1996). \textsl{A density-based algorithm for discovering clusters
			in large spatial databases with noise.} Proceedings of the Second International Conference on
		Knowledge Discovery and Data Mining (KDD-96). AAAI Press, Portland, Oregon, 226--231.	
		\bibitem{}
		Gower, J. C. (1971). \textsl{A general coefficient of similarity and some of its
			properties.} Biometrics. \textbf{27}, 623--637.
		\bibitem{}
		Handl, J., Knowles, K. \& Kell, D. (2005). \textsl{Computational cluster validation in post-genomic data analysis}. Bioinformatics. \textbf{21}, 3201--3212.
		\bibitem{}
		Hartigan, J. A. (1975). \textsl{Clustering Algorithms}. John Wiley \& Sons, New York, USA.
		\bibitem{}
		Hartigan, J. A. and Wong, M. A. (1979). \textsl{A K-means clustering algorithm}.
		Applied Statistics. \textbf{28}, 100--108.
		\bibitem{}
			Hahsler, M., Piekenbrock, M., Doran, D. (2019). \textsl{dbscan: Fast Density-Based Clustering with R}. Journal of Statistical Software. \textbf{91}, 1--30.
		\bibitem{}
		Jain, A. K. , Murty, M. N. and Flynn, P. J. (1999). \textsl{Data clustering: a review}. ACM
		Computing Surveys. \textbf{31}, 264--323.
		\bibitem{}
		Karney, C.F.F. (2013). \textsl{Algorithms for geodesics}, Journal of Geodesy, \textbf{87}, 43--55.
		\bibitem{}
		Kaufman, L. and Rousseeuw, P. J. (2005). \textsl{Finding Groups in Data: An Introduction to Cluster Analysis.} John Wiley and Sons, New Jersey.
		\bibitem{}
		MacQueen, J. (1967). \textsl{Some methods for classification and analysis of multivariate observations}. In Proceedings of the Fifth Berkeley Symposium on Mathematical Statistics and Probability, eds L. M. Le Cam \& J. Neyman, \textbf{1}, pp. 281--297. University of California Press, Berkeley, CA.
		\bibitem{}
		Marozzi, M. (2015). \textsl{Multivariate multidistance tests for high-dimensional low sample size case-control studies}. Statistics in Medicine,
		\textbf{34}, 1511--1526.
		\bibitem{}
		Matioli, L. C., Santos,  S. R., Kleina,  M. \& Leite, E. A. (2018). \textsl{A new algorithm for clustering based on kernel density estimation}. Journal of Applied Statistics. \textbf{45}, 347--366.
		\bibitem{}
		McLachlan, G. and Peel, D. (2000). \textsl{Finite Mixture Models}. John Wiley and Sons,
		New York.
		\bibitem{}
		Modak, S. (2019). \textsl{Uncovering astrophysical phenomena related to galaxies and other objects through statistical analysis.} Ph.D. Thesis, URL: http://hdl.handle.net/10603/314773 
		\bibitem{}
		Modak, S. (2021). \textsl{Distinction of groups of gamma-ray bursts in the BATSE catalog through fuzzy clustering}. Astronomy and Computing. \textbf{34}, Article id 100441, 1--7.
		\bibitem{}
		Modak, S. (2022a). \textsl{A new nonparametric interpoint distance-based measure for assessment of clustering}. Journal of Statistical Computation and Simulation. \textbf{9},	1062--1077.
		\bibitem{}
		Modak, S. (2022b). \textsl{A new measure for assessment of clustering based on kernel	density estimation}. Communications in Statistics -- Theory and Methods. In Press, Doi: 10.1080/03610926.2022.2032168	
		\bibitem{}
		Modak, S. \& Bandyopadhyay, U. (2019). \textsl{A new nonparametric test for
			two sample multivariate location problem with application to astronomy}. Journal of Statistical Theory and Applications. \textbf{18}, 136--146.
		\bibitem{}
		Modak, S., Chattopadhyay, A. K. \& Chattopadhyay, T. (2018). \textsl{Clustering of gamma-ray bursts through kernel principal component analysis}. Communications in Statistics -- Simulation and Computation. \textbf{47}, 1088--1102.
		\bibitem{}
		Modak, S., Chattopadhyay, T. \& Chattopadhyay, A. K. (2017). \textsl{Two
			phase formation of massive elliptical galaxies: study through cross--correlation including spatial effect.} Astrophysics and Space Science. \textbf{362}, Article id: 206, pages 1--10.
		\bibitem{}
		Modak, S., Chattopadhyay, T. \& Chattopadhyay, A. K. (2020). \textsl{Unsupervised classification of eclipsing binary light curves through k-medoids
			clustering}. Journal of Applied Statistics. \textbf{47}, 376--392.
		\bibitem{}
		Modak, S., Chattopadhyay, T. \& Chattopadhyay, A. K. (2022). \textsl{Clustering of eclipsing binary light curves through functional principal component analysis}. Astrophysics and Space Science. \textbf{ 367}, Article id: 19, pages 1--10
		\bibitem{}
		Nelsen, R. B. (2006). \textsl{An Introduction to Copulas, 2nd edition.} Springer Science+Business, New York.
		\bibitem{}
		Rousseeuw, P. J. (1987). \textsl{Silhouettes: A graphical aid to the interpretation and validation of cluster analysis.} Journal of Computational and Applied Mathematics. \textbf{20}, 53--65. 
		\bibitem{}	
		Ruspini, E. H.  (1970). \textsl{Numerical methods for fuzzy clustering.} Information Sciences. \textbf{2}, 319--350.
		\bibitem{}
		Sch\"{o}lkopf, B. and Smola, A. J. (2002). \textsl{Learning with Kernels: Support Vector Machines, Regularization, Optimization, and Beyond}. MIT Press. Cambridge.
		\bibitem{}
		Silverman, B. W. (1986), {\it Density Estimation for Statistics and Data Analysis}, Chapman
		and Hall, London.
		\bibitem{}
		Tarnopolski, M. (2019). \textsl{Analysis of the Duration--Hardness Ratio Plane of Gamma-Ray Bursts Using Skewed
			Distributions.} The Astrophysical Journal. \textbf{870}, 1--9, Article id: 105. 
		\bibitem{}
		T\'{o}th, B. G., R\'{a}cz, I. I. \& Horv\'{a}th, I. (2019). \textsl{Gaussian-mixture-model-based cluster analysis of gamma-ray bursts in
			the BATSE catalog}. Monthly Notices of the Royal Astronomical Society. \textbf{486}, 4823--4828.
		\bibitem{}
		Wand, M. P. and Jones, M. C. (1995), {\it Kernel Smoothing}, Chapman and Hall, New York.
	\end{thebibliography}
\end{document}